\documentclass[a4paper,11pt]{article}
\usepackage{pos}
\usepackage{amsmath}
\usepackage{simpler-wick}

\title{Baryon masses with C-periodic boundary conditions}

\author[a]{Anian Altherr}
\author[b]{Isabel Campos}
\author[a]{Roman Gruber}
\author[a]{Tim Harris}
\author[c]{Francesca Margari}
\author[a]{Marina Krstić Marinković}
\author[a]{Letizia Parato}
\author[d,e]{Agostino Patella}
\author*[b,d]{Sara Rosso}
\author[a]{Paola Tavella}

\affiliation[a]{Institut für Theoretische Physik, ETH Zürich, Wolfgang-Pauli-Str. 27, 8093 Zürich, Switzerland}
\affiliation[b]{Instituto de Física de Cantabria (IFCA) and Consejo Superior de Investigaciones Cientificas (CSIC), Avda. de Los Castros s/n, 39005 Santander, Spain}
\affiliation[c]{Dipartimento di Fisica, Università di Roma “Tor Vergata” and INFN, Sezione di Roma “Tor Vergata”, Via della Ricerca Scientifica 1, I-00133 Roma, Italy}
\affiliation[d]{Humboldt Universität zu Berlin, Institut für Physik and IRIS Adlershof, Zum Großen Windkanal 6, 12489
Berlin, Germany}
\affiliation[e]{DESY, Platanenallee 6, D-15738 Zeuthen, Germany}

\emailAdd{rosso@ifca.unican.es}
\abstract{Isospin-breaking corrections pose a significant challenge to lattice simulations, both because of the splitting between the up and down quark masses and, in particular, the need to include QED effects. The RC* collaboration has developed the openQxD code, based on openQCD, which enables fully dynamical QCD+QED simulations through the implementation of C-periodic boundary conditions.
We use this code to measure baryon masses, with a special focus on the $\Omega^-$ baryon mass, whose precise determination is especially important since it has been used to set the scale of lattice simulations.
Due to the use of C-periodic boundary conditions, the two-point function of the $\Omega^-$ baryon gets additional partially connected contributions, which vanish in the infinite-volume limit and which we are computing for the first time.
We will present preliminary results for baryon masses obtained on QCD ensembles with C-periodic boundary conditions, at an unphysical pion mass of approximately 400 MeV.}

\FullConference{The 42nd International Symposium on Lattice Field Theory (LATTICE2025)\\
2-8 November 2025\\
Tata Institute of Fundamental Research, Mumbai, India\\}

\begin{document}
\maketitle

\section{Isospin breaking and C-periodic boundary conditions}
\begin{figure}
    \centering
    \includegraphics[width=0.7\linewidth]{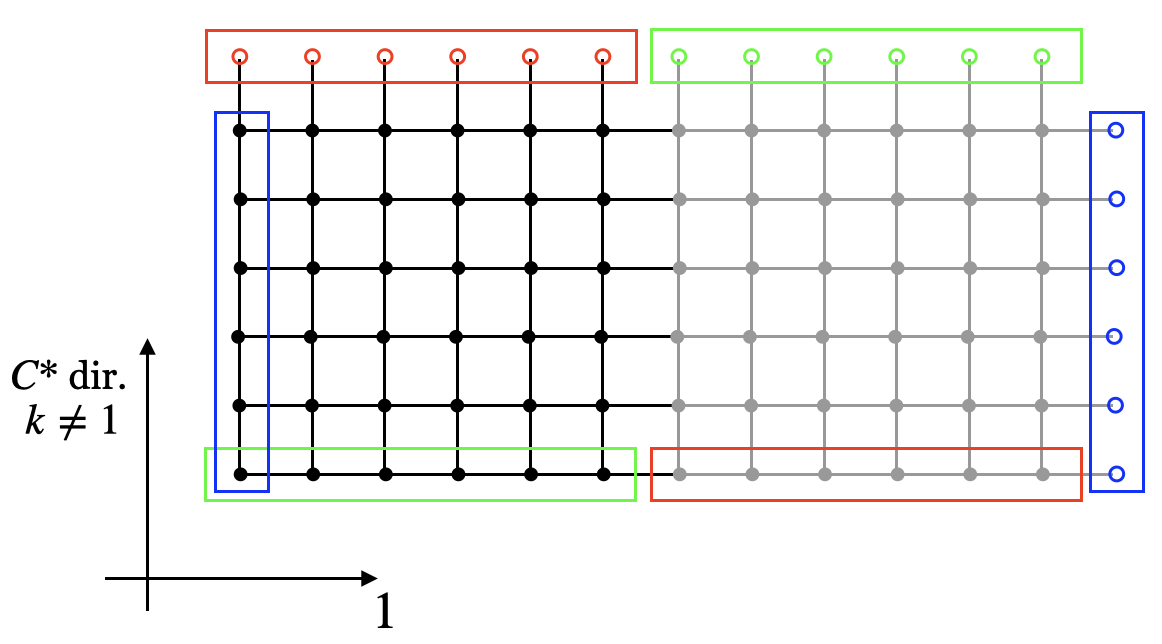}
    \caption{2-dimensional section of a 4-dimensional lattice representing the orbifold construction used to implement C-periodic boundary conditions. Figure taken from \cite{open.code}}.
    \label{fig:bound}
\end{figure}

Lattice Quantum Chromodynamics (QCD) simulations have reached a stage where precision is essential, with the goal of achieving results at the percent or sub-percent level on many hadronic observables. At this precision, the theory does not present exact isospin symmetry, with isospin defined as rotations of the light quarks in flavour space. For this reason, it has become crucial to include the computation of isospin breaking corrections, coming both from the difference in mass and in charge of the light quarks. To account for the different electric charges, we need to consider the coupling to Quantum Electrodynamics (QED).\\
The definition of QED in finite volume is made difficult by the long range of the electromagnetic interaction, as we can see by looking at the total charge in a finite volume with periodic boundary conditions in space, using the Gauss law:
\begin{equation}
    Q = \int_0^L d^3x \rho(x) = \int_0^L dx^3 \nabla \cdot E(x) = 0
    \label{eq:gauss}
\end{equation}
Due to the electric field being periodic at the boundary of the lattice, the result is always zero. Consequently, it is impossible, for example, to measure the mass of a charged particle because no interpolating operator can have a non-zero overlap with a charged state.\\
Several definitions of QED in finite volume have been proposed, and we rely on the so-called QED$_C$, a formulation that preserves locality, gauge invariance and translational invariance, implemented by the RC* collaboration in the \texttt{openQxD} code \cite{openqxd}, \cite{open.code}, as an extension of \texttt{openQCD-1.6} \cite{openqcd}.
This definition of QED involves a particular choice of boundary conditions in space, C-periodic boundary conditions, that consist in the charge conjugation of the fields of the theory when crossing the boundaries of the lattice \cite{char.had}.
\begin{equation}
    \begin{split}
    &A_\mu \left( x + \hat{L}_i \right) = A^C_\mu (x) = -A_\mu (x) \\
    &U_{\rho}(x + \hat{L}_i) = U_{\rho}^C(x)= U_{\rho}(x)^*\\
    &q_f \left( x + \hat{L}_i \right) = q_f^C (x) = C^{-1} \bar{q}_f^T (x)\\
    &\bar{q}_f \left( x + \hat{L}_i \right) = \bar{q}_f^C (x) = -q_f^T (x) C
\end{split}
\label{eq:Cbound}
\end{equation}
$A_\mu$ is the electromagnetic gauge field, $U_{\rho}$ are the colour gauge fields, $q_f$ is the quark field of flavour $f$, $C=i\gamma_0\gamma_2$ is the charge conjugation matrix and $L_i$ is the extension of the lattice in the space direction $i$.\\
These boundary conditions solve the issue of zero charge on the lattice, as we can easily see repeating the reasoning of equation (\ref{eq:gauss}); in this case the difference of the electric field at the boundaries can be different from zero because the field is anti-periodic. \\
In the \texttt{openQxD} code, these boundary conditions are implemented through an orbifold construction, shown in figure (\ref{fig:bound}), where we have a 2-D section of the 4-dimensional lattice, with the first space direction on the horizontal axis and another space direction on the vertical axis. In this construction, the physical lattice is doubled along the first space direction with an exact replica, the mirror lattice, where the fields are defined as the charge conjugated of the ones in the physical lattice. Shifting a field of $L_i$ means moving from the physical to the mirror lattice and vice versa. In this way the boundary conditions are implemented by definition and we also obtain some periodicity conditions in the extended lattice, as shown by the domains of different colours in the figure.\\
C-periodic boundary conditions modify the structure of the quark propagators, leading to non-zero contributions not only for the quark–antiquark propagator, but also for the quark–quark and antiquark–antiquark propagators. These additional terms give an element of the Dirac operator where one of the two points has been shifted to the mirror lattice, as shown in equations (\ref{eq:wick}), where for simplicity we perform all the shifts of $L$ in the first space direction.
\begin{subequations}
    \begin{equation}
        \langle \wick{\c1{q}(x) \, \c1{\overline q}(y)} \rangle = D^{-1}(x, y)
    \end{equation}
    \begin{equation}
        \Big\langle \wick{\c1{q}(x) \, \c1{q}^{T}(y)} \rangle = - D^{-1}(x, y +L\hat{1}) \, C  
    \label{eq:wickq}
    \end{equation}
    \begin{equation}
        \langle \wick{\c1{\overline q}^{T}(x) \, \c1{\overline q}(y)} \rangle = C \, D^{-1}(x +L\hat{1}, y)
    \label{eq:wick.qbar}
    \end{equation}
    \label{eq:wick}
\end{subequations}
These additional propagators determine the presence of some partially connected contributions in the baryonic two-point functions with C-periodic boundary conditions, as we show in the following section.
\section{Two-point function of the $\Omega^-$ baryon}
Baryons are hadrons with three valence quarks, for this reason they can be classified according to the fundamental representation of the approximate $SU(3)$ flavour group. The lowest dimensional representations are the octet, an 8-dimensional representation including the proton and neutron, and the decuplet, an 10-dimensional representation including the $\Omega^-$, a baryon with three strange quarks as valence quarks, $3/2$ spin and positive parity. In the following we will keep a special focus on the $\Omega^-$, which is especially interesting since its mass has been used to set the scale of lattice simulations (\cite{BMW}, \cite{ukqcd}, \cite{scale.set}).\\
We define interpolating operators for the $\Omega^-$ containing three strange quark fields $\psi$: 
\begin{subequations}
    \begin{equation}
        v_{\Omega^-}^{m;d}(x) = \sum_{\substack{abc\\ABC}} W^{d;m}_{abc} \epsilon^{ABC} \psi^C_c(x) \psi^A_a(x) \psi^B_b(x)
    \end{equation}
    \begin{equation}
                \overline{v}_{\Omega^-}^{m;d}(x) = \sum_{\substack{abc\\ABC}} \overline W^{d;m}_{abc} \epsilon^{ABC} \overline{\psi}^B_b(x) \overline{\psi}^A_a(x) \overline{\psi}^C_c(x)
    \end{equation}
\end{subequations}
where
\begin{subequations}
    \begin{equation}
        W^{d;m}_{abc} = P^{ml}_{dc} \Gamma^l_{ab} + P^{ml}_{db} \Gamma^l_{ac} + P^{ml}_{da} \Gamma^l_{cb}  
    \end{equation}
    \begin{equation}
         \overline W^{d;m}_{abc} = P^{ml}_{cd} \Gamma^l_{ab} + P^{ml}_{bd} \Gamma^l_{ac} + P^{ml}_{ad} \Gamma^l_{cb}
    \end{equation}
    \begin{equation}
        P^{ml} = [ \delta^{ml} Id_{4\times4} - \frac{1}{3} \gamma^{m} \gamma^{l} ] 
    \end{equation}
    \begin{equation}
        \Gamma^{l} = C\gamma^{l}
    \end{equation}
\end{subequations}
small letters $a-d$ are Dirac indices, small letters $m,l$ are space indices, capital letters are colour indices and $P^{ml} $ is the projector to $3/2$ spin. The three terms in the definition of the $W$ and $\overline{W}$ tensors are introduced to symmetrize the interpolating operators in the exchanges of the three strange quark fields.\\
With the interpolating operators we can construct the two-point function projected to zero momentum and positive parity
\begin{equation}
    C_{\Omega^-}(x_0) = \sum_{\substack{dd'\\m}} \sum_{\textbf{x}} P^+_{dd'} \;v_{\Omega^-}^{m;d}(x) \overline{v}_{\Omega^-}^{m;d'}(0)
\end{equation}
where $P^+ = (1+\gamma^0)/2$ is the projector to positive parity.\\
We insert the definitions of the interpolating operators into the two-point function and perform all possible Wick contractions of the quark and antiquark fields, finding that two different types of contraction arise.\\
The first kind, in equation (\ref{eq:3q2p}), is the standard result that is obtained for the two-point function in the case of periodic boundary conditions in space. All three Wick contractions are performed between a quark and an antiquark field, resulting in three quark lines connecting the two space-time points; for this reason, we refer to these contractions as 3-quark (3-q) connected.
\begin{equation}
\begin{split}
    C_{\Omega^-}^{3-q}(x_0)=  \sum_{\substack{d'd\\m}}  \sum_{\substack{a bc\\a'b'c'}} \sum_{\substack{\\ABC\\A'B'C'}} \sum_{\textbf{x}}  \Big[\overline W^{d';m}_{a'b'c'} P^+_{dd'}  W^{d;m}_{abc} \epsilon^{ABC} \epsilon^{A'B'C'}  \wick{ \c1{\psi}^A_a(x)\c1{\overline{\psi}}^{A'}_{a'}(0) } \wick{ \c1{\psi}^B_b(x)\c1{\overline{\psi}}^{B'}_{b'}(0) }\\ \wick{ \c1{\psi}^C_c(x)\c1{\overline{\psi}}^{C'}_{c'}(0) } \Big]
    \label{eq:3q2p}
    \end{split}
\end{equation}
The second kind, in equation (\ref{eq:1q2p}), is an artefact of C-periodic boundary conditions, since it is obtained using the modified quark propagators in equation (\ref{eq:wick}), with one quark-antiquark propagator connecting the two space-time points, one quark-quark propagator at the sink point and one antiquark-antiquark propagator at the source point. For this reason, these contributions are partially connected and we refer to them as 1-quark (1-q) connected. We expect these contributions to vanish exponentially in the infinite volume limit, as detailed in \cite{char.had}.
\begin{equation}
    \begin{split}
        C_{\Omega^-}^{1-q}(x_0)= \sum_{\substack{d'd\\m}}  \sum_{\substack{a bc\\a'b'c'}}\sum_{\substack{\\ABC\\A'B'C'}} \sum_{\textbf{x}} \Big[ \overline W^{d';m}_{a'b'c'} P^+_{dd'}  W^{d;m}_{abc} \epsilon^{ABC} \epsilon^{A'B'C'}  \wick{\c1{\overline{\psi}}^{A'}_{a'}(0) \c1{\overline{\psi}}^{B'}_{b'}(0)} \wick{  \c1 \psi_{c}^{C}(x)\c1{\overline{\psi}}^{C'}_{c'}(0) } \\\wick{\c1{\psi}^B_b(x)\c1{\psi}^A_a(x) } \Big]
    \end{split}
    \label{eq:1q2p}
\end{equation}
\section{Numerical results}
In this section, we will present some preliminary results for the computations of the two kinds of contributions to the baryonic two-point functions. Other computations of baryon masses have been  performed by our collaboration and discussed in \cite{first.res}. We want to highlight some relevant differences with those measurements, first of all the absence in the previous computation of the 1-q connected contributions, that we are computing for the first time. Second, considering the 3-q connected contributions only, we aim to increase the number of point sources used in the computation by one order of magnitude, going from $O(10)$ sources to $O(100)$, as we will detail in the following.\\
We will show results for two ensembles generated by our collaboration, \texttt{B400a00b324} and\\  \texttt{A400a00b324}, whose parameters are shown in table (\ref{tab:ensemb}). They are two QCD-only ensembles with C-periodic boundary conditions, with three degenerate light quarks and the charm quark, $O(a)$ improved Wilson fermions and an unphysical pion mass of around 400 MeV. For $O(a)$ improvement we adopt $c_{sw}^{SU(3)} = 2.18859$, determined in isosymmetric QCD \cite{fritzsch}. The scale setting is performed via the gradient-flow observable $t_0$, 
using the central value reported for $N_f = 3$ by the CLS collaboration \cite{hollwieser}, 
$\sqrt{8t_0} = 0.415\,\mathrm{fm}$. The two ensembles differ for the volume of the lattice, bigger in the \texttt{B400a00b324} ensemble. The choice of QCD-only ensembles is motivated by the fact that the computations are less demanding in this case, for all the baryons belonging to the octet and all belonging to the decuplet are degenerate and we can compute two masses instead of five. We plan to analyze a QCD+QED ensemble during this year, for which computational resources have already been allocated.
\begin{table}[]
    \centering
    \begin{tabular}{|c|c|c|c|}
        \hline
         Ensemble & $m_{\pi^0}$ [MeV] & $a$[fm] & V \\
         \hline
          \texttt{B400a00b324} & 401.9(1.4) & 0.05400(14) & $80\times48^3$\\
          \texttt{A400a00b324} & 398.5(4.7) & 0.05393(24) & $64\times32^3$ \\
         \hline
    \end{tabular}
    \caption{Ensembles parameters as detailed in \cite{first.res}}
    \label{tab:ensemb}
\end{table}

\subsection{3-q connected contributions}
\begin{figure}
    \centering
    \includegraphics[width=1\linewidth]{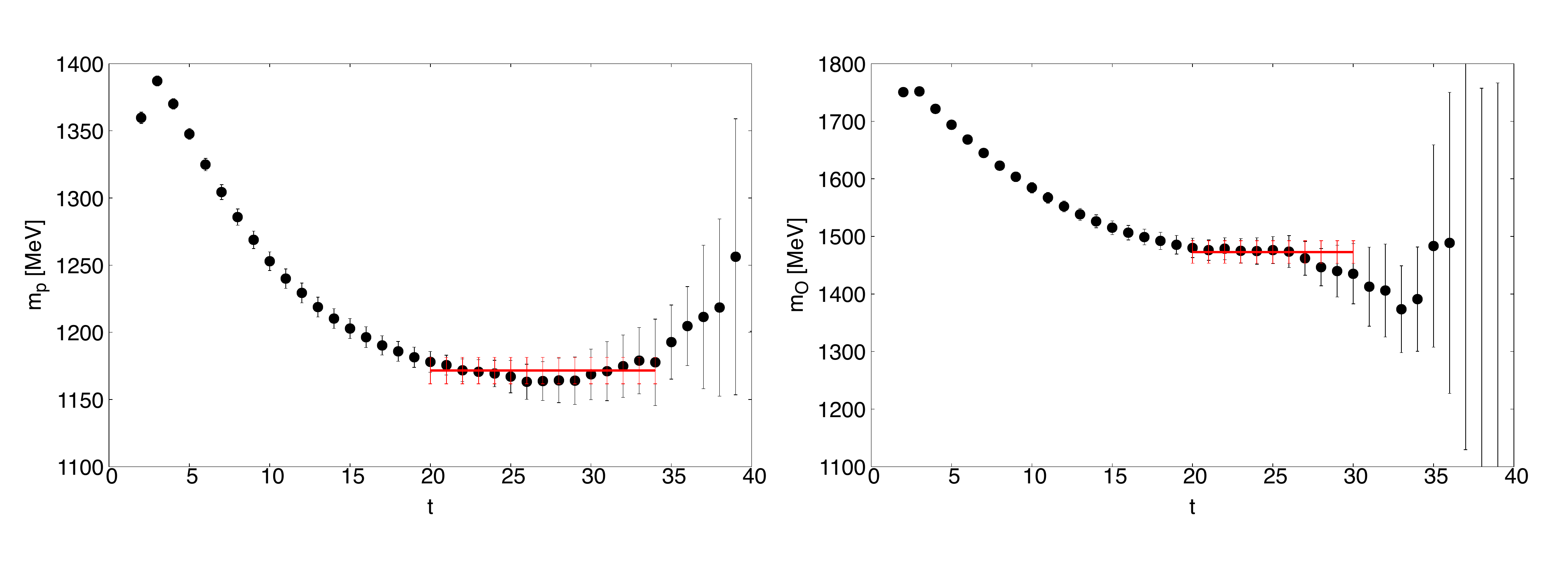}
    \caption{Proton and $\Omega^-$ effective masses together with the selected plateaux and the fits to a constant for the ensemble \texttt{B400a00b324}. Values in MeV are obtained by using the reference value $(8t_0 )^{1/2} =$ 0.415 fm.}
    \label{fig:B400}
\end{figure}

\begin{figure}
    \centering
    \includegraphics[width=\linewidth]{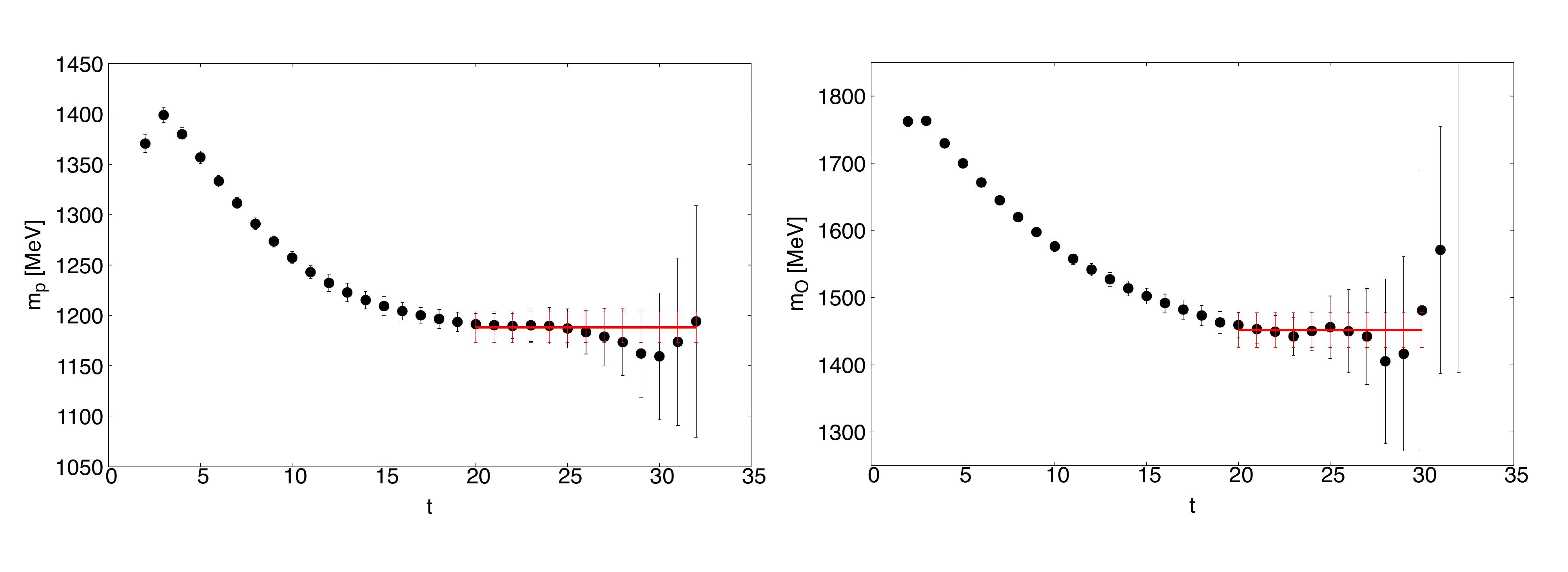}
    \caption{Proton and $\Omega^-$ effective masses together with the selected plateaux and the fits to a constant for the ensemble \texttt{A400a00b324}. Values in MeV are obtained by using the reference value $(8t_0 )^{1/2} =$ 0.415 fm.}
    \label{fig:A400}
\end{figure}

In this section we show the results for the computation of the 3-q connected contributions to baryonic two-point functions, as in equation (\ref{eq:3q2p}). Those involve only quark–antiquark propagators; therefore, they contain inverses of the Dirac operator at a fixed source point and can be computed using point sources. As explained in the previous section, for the two ensembles considered we measure two baryon masses, the proton and the $\Omega^-$ mass, considering that all the baryons in the octet have the same mass of the proton and all those in the decuplet the same as the $\Omega^-$.\\
We use smearing to better the superposition of the interpolating operators with the desired physical state. We smear the gauge fields using the gradient-flow smearing \cite{grad.flow} and use the smeared fields as input for Gaussian fermion smearing \cite{smear}:
\begin{equation}
    \psi_{\text{smeared}}(x) = \left( \frac{1 + \kappa H_t(x, y)}{1 + 6\kappa}\right)^N \psi(y)
\end{equation}
where $H_t(x, y)$ is the spatial hopping operator 
\begin{equation}
    H_t(x, y) = \sum_{j=1}^{3} \left[ V_t(x, j)\, \delta(x + \hat{j}, y) + V_t(x - \hat{j}, j)^\dagger\, \delta(x - \hat{j}, y) \right]
\end{equation}
and $V_t$ the gauge-smeared links. The strength $\kappa$ and the number of smearing steps $N$ were chosen according to the best parameters for the results presented in \cite{first.res}. The smearing steps can be organized in different smearing levels and at each level the results are printed. This provides the necessary input for a GEVP analysis \cite{gevp}, which will be carried out after results from all sources have been obtained.\\
We show in figure (\ref{fig:B400}) the results for the proton and $\Omega^-$ effective masses with the selected plateaux and fit to constant for the ensemble \texttt{B400a00b324}. These results were obtained using 60 point sources to invert the Dirac operator and maximum smearing at the source and no smearing at the sink; we observe that this combination minimizes the noise, as was also observed in the analysis of the results shown in \cite{first.res}.\\
In figure (\ref{fig:A400}) are the proton and $\Omega^-$ effective masses with the selected plateaux and fit to constant for the ensemble \texttt{A400a00b324}. In this case 50 point sources were used to invert the Dirac operator and we use the same combination of smearing levels.
The results for the masses are also shown in table (\ref{tab:masses}) for the two ensembles.\\
We observe a reduction of noise in the results, compared to those in \cite{first.res} that were obtained using either 4 or 8 point sources. The reduction in noise, along with the inclusion of the large-volume ensemble \texttt{B400a00b324}, allows us to identify longer plateaux and obtain more reliable mass values. 

\begin{table}[]
    \centering
    \begin{tabular}{|c|c|c|}
        \hline
        Ensemble  & $m_p$[MeV] & $m_{\Omega^-}$[MeV] \\
        \hline
         \texttt{B400a00b324}& 1170(10) & 1470(20)\\
         \texttt{A400a00b324}& 1190(15) & 1450(25) \\
         \hline
    \end{tabular}
    \caption{Proton and $\Omega^-$ masses for the two ensembles considered.}
    \label{tab:masses}
\end{table}
\subsection{1-q connected contributions}
\begin{figure}
    \centering
    \includegraphics[width=1\linewidth]{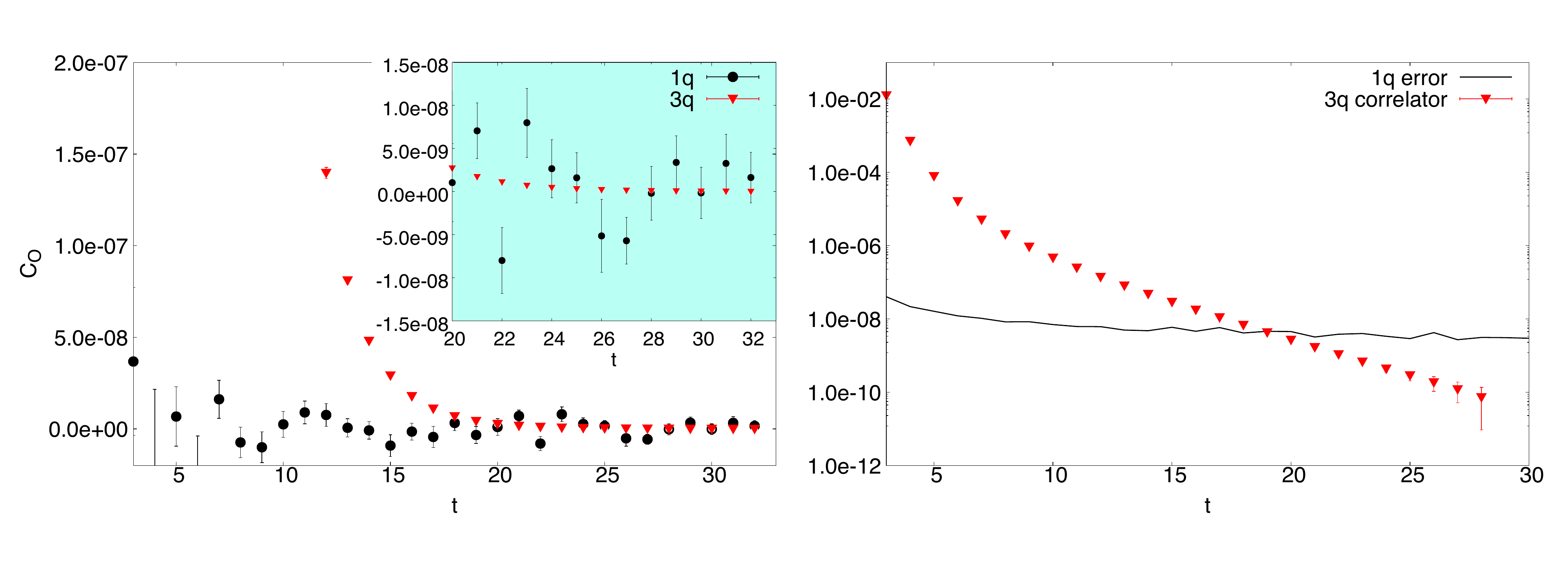}
    \caption{Results for the $\Omega^-$ 2-point function with no smearing. In the left plot are the 3-q connected contributions, in red, and the 1-q connected, in black, with a zoom of the high $t$ region. In the right plot is a comparison of the 3-q correlator with the absolute error on the 1-q correlator, in a logarithmic scale for the y axis.}
    \label{fig:1q_sm0}
\end{figure}

\begin{figure}
    \centering
    \includegraphics[width=1\linewidth]{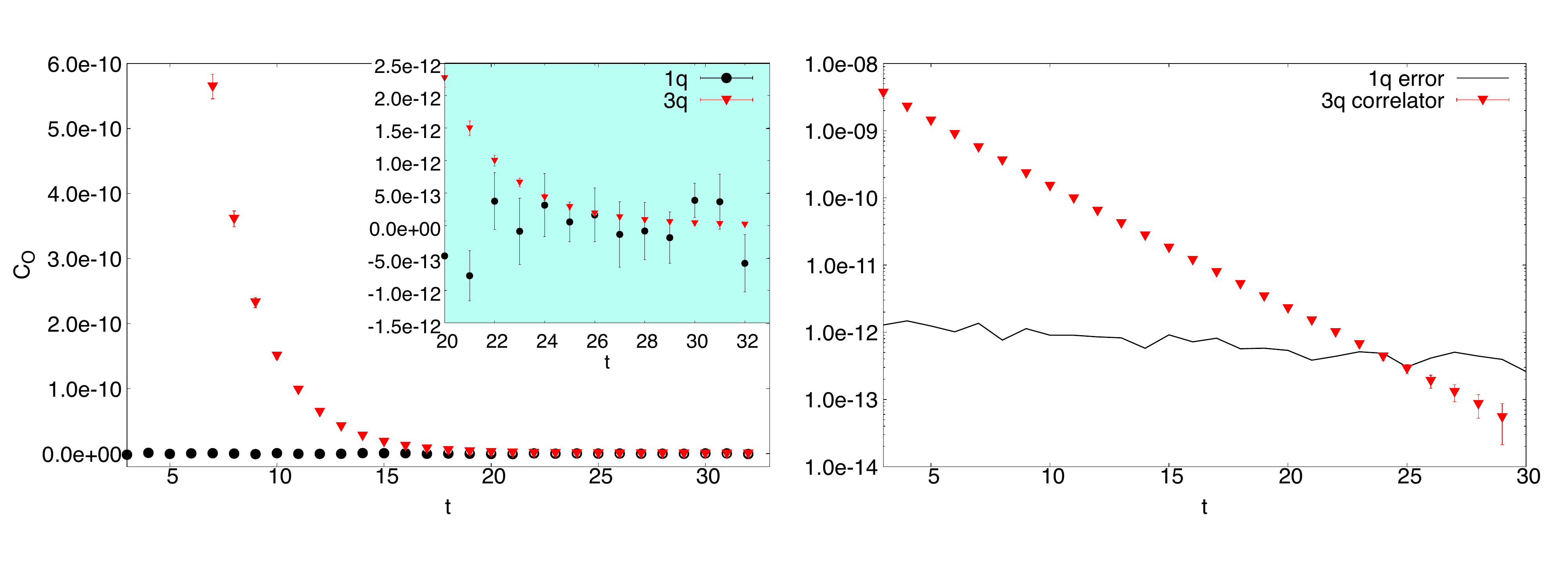}
    \caption{Results for the $\Omega^-$ 2-point function with maximum smearing on the source and none on the sink. In the left plot are the 3-q connected contributions, in red, and the 1-q connected, in black, with a zoom of the high $t$ region. In the right plot is a comparison of the 3-q correlator with the absolute error on the 1-q correlator, in a logarithmic scale for the y axis. }
    \label{fig:1_q.sm6}
\end{figure}

In this section, we focus on the $\Omega^-$, detail the implementation of the 1-q connected contributions and show some preliminary results for this quantity measured on the \texttt{A400a00b324} ensemble.\\
We can rewrite the two-point function in equation (\ref{eq:1q2p}) substituting the results for the modified quark propagators in equation (\ref{eq:wick}). We obtain:
\begin{equation}
\begin{split}
    C_{\Omega^-}^{1-q}(x_0) =-  \sum_{\substack{d'd\\m}}  \sum_{\substack{\alpha abc\\\alpha'a'b'c'}}\sum_{\substack{\\ABC\\A'B'C'}} \sum_{\textbf{x}} \Big[ \overline W^{d';m}_{a'b'c'} P^+_{dd'}  W^{d;m}_{abc} \epsilon^{ABC} \epsilon^{A'B'C'}
    C_{a'\alpha'} \;D^{-1}(L\hat{1},0)^{A'B'}_{\alpha' b'} \\ D^{-1}(x,0)^{CC'}_{cc'}  \; \;D^{-1}(x,x+L\hat{1})^{BA}_{b\alpha}  \; C_{\alpha a} \Big]
    \end{split}
    \label{eq:1q}
\end{equation}
The two inverses of the Dirac operator at fixed source point, fixed to the origin in equation (\ref{eq:1q}) for simplicity, can be computed using point sources. 
A different treatment is required for the last inversion in equation (\ref{eq:1q}), which is an all-to-all propagator. This is computed using stochastic sources 
$\chi^{(n)}$, constructed by independently generating 12 random numbers at each lattice site, one for each colour–Dirac component. We employ the identity relation:
\begin{equation}
   \left< \frac{1}{N_s} \sum_n \chi(x)^{(n)\dagger A}_a \chi(y) ^{(n)B}_b \right>= \delta_{AB}\delta_{ab}\delta_{xy} 
\end{equation}
to obtain a stochastic estimator for the 1-q connected contributions to the $\Omega^-$ two-point function:
\begin{equation}
    \begin{split}
    C_{\Omega^-}^{1-q}(x_0) =-  \sum_{\substack{d'd\\m}}  \sum_{\substack{\alpha abc\\\alpha'a'b'c'}} \sum_{\substack{\\ABC\\A'B'C'}} \sum_{\textbf{x}} \Big[ \overline W^{d';m}_{a'b'c'} P^+_{dd'}  W^{d;m}_{abc} \epsilon^{ABC} \epsilon^{A'B'C'}
    C_{a'\alpha'} \;D^{-1}(L\hat{1},0)^{A'B'}_{\alpha' b'} \\ D^{-1}(x,0)^{CC'}_{cc'}
    \frac{1}{N_s} \sum_n \left[D^{-1}\chi^{(n)}\right](x)  _{b}^{B}\;\; \chi^{\dagger(n)}(x+L\hat{1})^{A}_{a} \Big]
    \end{split}
\end{equation}
We implemented the code for these measurements as a module of the \texttt{openQxD} code, completing the work previously presented in \cite{lattice24}. The code was tested by checking gauge and translational invariance of the two-point function and the tree-level result.\\
We measured these contributions for 40 decorrelated configurations, spaced 50, of the \texttt{A400a00b324} ensemble, using 10 point and 10 stochastic sources for the two inversions of the Dirac operator. We apply smearing to the source and sink, with the same smearing operators and input parameters as for the 3-q connected contributions. We know that the 1-q connected contributions vanish in the infinite volume limit and we want to compare the relative size of the two kinds of contributions at finite volume, where we expect the 1-q connected contributions to be small compared to the 3-q connected ones. At the same time, since the 1-q connected contributions involve a stochastic estimator and inverses of the Dirac operator spaced at the lattice extension $L$, we expect these to be noise-dominated.\\
We compare the results for the two kinds of contributions for the same configurations, with the 3-q connected obtained using 10 point sources to invert the Dirac operator.  The results without smearing are shown in figure (\ref{fig:1q_sm0}). Noise makes it difficult to compare the two contributions in the large-$t$ region, since the error on the 1-q connected term is approximately two orders of magnitude larger than that of the 3-q connected term.
In the right plot in figure (\ref{fig:1q_sm0}) is a comparison, in a logarithmic scale on the y axis, of the 3-q connected correlator and the absolute error on the 1-q connected contributions. The intersection marks the time separation for which the error on the 1-q contributions becomes dominant. We see that for no smearing the error on the 1-q connected is already dominant at $t=19$.\\
In figure (\ref{fig:1_q.sm6}) are the results for maximum smearing at the source and none at the sink, as we found it to be the best combination for the 3-q connected contributions. We see a significant decrease in the noise and the two kinds of contributions becoming of the same order at high $t$. Looking at the left plot in figure (\ref{fig:1_q.sm6}) we see that the the error on the 1-q connected is dominant starting at $t=24$, which also marks a significant improvement compared with no smearing.

\section{Conclusions and outlook}
We provide computations of baryon masses for two QCD ensembles with C-periodic boundary conditions, a fundamental step toward including isospin corrections, despite the well-known difficulties in computing baryon masses due to the signal-to-noise problem. We computed the 3-q connected contributions, which constitute the bulk of the computation, with a high number of point sources used to invert the Dirac operator, proving a reduction of the noise with respect to previous results. We plan to increase the number of sources to 100 for the large-volume ensemble and 150 for the small-volume ensemble. We obtained results for different combinations of smearing levels, which we plan to use in a GEVP analysis at the end of the allocation. We will also extend the analysis to a QCD+QED ensemble.\\
We also computed for the first time the 1-q connected contributions to the $\Omega^-$ two-point function, a finite volume effect that we expect to vanish exponentially in the infinite volume limit. We observed the importance of smearing the operators to better the signal and make the comparison with the 3-q connected contributions possible. We are investigating the choice of using stochastic sources to compute the all-to-all propagator, working to find a definition of the estimator which has a variance which vanishes with the lattice size $L$, similarly to the expected signal.  

\acknowledgments
This research project was made possible through the access granted by the Galician Supercomputing Center (CESGA) to its supercomputing infrastructure. 
We acknowledge access to Lise at NHR@ZIB as part of the NHR infrastructure (project bep00138) .

\end{document}